\documentclass[english,10pt,letterpaper,twocolumn,floatfix,allowfontchangeintitle,accepted=2022-01-31]{quantumarticle}
\pdfoutput=1
\usepackage[utf8]{inputenc}
\usepackage{amsmath}
\usepackage{amssymb}
\usepackage{bm}
\usepackage{bbm}
\usepackage{epsfig}
\usepackage{grffile}
\usepackage{graphics}
\usepackage[export]{adjustbox}

\usepackage[numbers,sort&compress]{natbib}
\usepackage{scalefnt}

\usepackage[usenames,dvipsnames]{xcolor}
\definecolor{dblue}{rgb}{0,0.1,.6}

\usepackage[colorlinks=true,citecolor=dblue,linkcolor=dblue,urlcolor=dblue]{hyperref}
\usepackage[all]{hypcap}

\newcommand{\Tr}{\operatorname{Tr}}
\newcommand{\bra}{\langle}
\newcommand{\ket}{\rangle}
\newcommand{\mc}[1]{\mathcal{#1}}
\newcommand{\dm}{\varrho}
\newcommand{\td}{\text{td}}
\newcommand{\eff}{\text{eff}}
\newcommand{\cft}{\text{CFT}}
\newcommand{\tI}{\text{tI}}

\newcommand{\duke} {Department of Physics, Duke University, Durham, North Carolina 27708, USA}

\newcommand{\Title} {Eigenstate entanglement scaling for critical interacting spin chains}

\newcommand{\Authors}
{
\author{Qiang Miao}
\affiliation{\duke\vspace{-0.25em}}
\author{Thomas Barthel\vspace{-0.5em}}
\affiliation{\duke\vspace{-0.25em}}
}
\newcommand{\Date} {December 12, 2021}

\begin{document}

\title{\scalefont{0.92}\Title}
\Authors
\date{\Date}

\begin{abstract}
With increasing subsystem size and energy, bipartite entanglement entropies of energy eigenstates cross over from the groundstate scaling to a volume law. In previous work, we pointed out that, when strong or weak eigenstate thermalization (ETH) applies, the entanglement entropies of all or, respectively, almost all eigenstates follow a single crossover function. The crossover functions are determined by the subsystem entropy of thermal states and assume universal scaling forms in quantum-critical regimes. This was demonstrated by field-theoretical arguments and the analysis of large systems of non-interacting fermions and bosons. Here, we substantiate such scaling properties for integrable and non-integrable interacting spin-1/2 chains at criticality using exact diagonalization. In particular, we analyze XXZ and transverse-field Ising models with and without next-nearest-neighbor interactions. Indeed, the crossover of thermal subsystem entropies can be described by a universal scaling function following from conformal field theory. Furthermore, we analyze the validity of ETH for entanglement in these models. Even for the relatively small system sizes that can be simulated, the distributions of eigenstate entanglement entropies are sharply peaked around the subsystem entropies of the corresponding thermal ensembles.
\end{abstract}

\maketitle
\hypersetup{ pdftitle = {} }

\section{Introduction}
Entanglement is a fundamental feature of quantum matter with far-reaching consequences for its macroscopic properties, complexity, and technological potential. It lies at the heart of modern physics and pervades various research fields \cite{Amico2008-80,Horodecki2009-81,Laflorencie2016-646}. Here, we focus on bipartite entanglement entropies
\begin{equation}\label{eq:S}
	S = -\Tr \dm_\mc{A} \ln{\dm_\mc{A}},\quad\text{where } \dm_\mc{A} = \Tr_\mc{B}|E\ket\bra E|
\end{equation}
is the reduced density matrix for an energy eigenstate $|E\ket$ of a bipartite system $\mc{H}_\mc{A}\otimes\mc{H}_\mc{B}$. In the following, we consider compact subsystems $\mc{A}$ and investigate their entanglement with the complement $\mc{B}$ as a function of energy and linear subsystem size $\ell$. The latter is defined such that $\operatorname{vol}\mc{A}=:\ell^d$, where $d$ is the number of spatial dimensions.

The scaling of entanglement entropies in groundstates has been studied extensively \cite{Eisert2008,Latorre2009,Laflorencie2016-646}. For the typical condensed matter systems in the thermodynamic limit, groundstate entanglement either obeys an area law $S_\text{gs}\propto\ell^{d-1}$ \cite{Srednicki1993,Callan1994-333,Latorre2004,Plenio2005,Cramer2006-73,Hastings2007-76,Brandao2013-9,Cho2018-8} or a log-area law $S_\text{gs}\propto\ell^{d-1}\ln{\ell}$ \cite{Srednicki1993,Callan1994-333,Holzhey1994-424,Vidal2003-7,Jin2004-116,Calabrese2004,Zhou2005-12,Wolf2005,Gioev2005,Barthel2006-74,Li2006,Lai2013-111,Murciano2020-8}.
In contrast, the entanglement entropy of random pure states is generally extensive, i.e., obeys a volume law $S\propto\ell^d$, which is a consequence of quantum typicality~\cite{Popescu2006-2,Goldstein2006-96,Gemmer2004}. Recent studies have shown the volume law of entanglement for broad classes of highly excited states \cite{Alba2009-10,Ares2014-47,Storms2014-89,Moelter2014-10,Keating2015-338} and for the  average eigenstate entanglement \cite{Vdimar2017-119,Vidmar2017-119b,Vidmar2018-121,Lu2019-99,Huang2019-938,LeBlond2019-100,Lydzba2020_06}.

In previous work, we argued that, based on the eigenstate thermalization hypothesis (ETH), the crossover of eigenstate entanglement entropy from the groundstate scaling at low energies and small $\ell$ to a volume law at larger energies or $\ell$ is captured by a single crossover function which assumes a universal scaling form in quantum-critical regimes \cite{Miao2019_05}. A system obeys ETH if all eigenstates (strong ETH) or almost all eigenstates (weak ETH) look locally thermal, i.e., if the expectation values of local observables $\bra E|O|E\ket$ are, in the thermodynamic limit, indistinguishable from expectation values $\Tr(\dm_\td O)$ of corresponding thermodynamical ensembles \cite{Deutsch1991-43,Srednicki1994-50,Rigol2008-452,Biroli2010-105,Beugeling2014-89,Kim2014-90,Alba2015-91,Lai2015-91,Dymarsky2018-97,Deutsch2018-81,Yoshizawa2018-120}. In particular, $\dm_\td=\dm_\td(E)$ could be chosen as the (global) microcanonical ensemble for energy $E$ or, due to the equivalence of ensembles in the thermodynamic limit, as any other thermodynamical ensemble with the same energy density. If this ETH equivalence holds for all local observables in subsystem $\mc{A}$, it also holds for subsystem density matrices $\dm_\mc{A}$ obtained from $|E\ket\bra E|$ in the sense that $\|\dm_\mc{A}-\Tr_\mc{B}\dm_\td\|_1$ is small. Thus, ETH also holds for their subsystem entropies. Now, the long-range physics of critical condensed matter systems in equilibrium can usually be described by field theories, and their subsystem entropies are then captured by scaling functions that depend on dimensionless parameters like energy ratios. So, in the presence of ETH, the entanglement entropies of (almost) all energy eigenstates are characterized by such a scaling function.

In Ref.~\cite{Miao2019_05}, we gave the eigenstate entanglement scaling functions in analytical form for critical one-dimensional (1d) systems based on conformal field theory and for $d$-dimensional fermionic systems with a Fermi surface. Numerically, we demonstrated the scaling behavior and weak ETH for the entanglement in systems of non-interacting fermions in $d=1,2,3$ \cite{Miao2019_05}, and for the harmonic lattice model (free scalar field theories) in $d=1,2$ \cite{Barthel2019_12}. Simulations for such non-interacting systems are efficient, and the results are basically free of finite-size effects.

In this paper, we probe and confirm the described scaling properties and the applicability of ETH for \emph{interacting} spin-$1/2$ chains in integrable and non-integrable regimes. Although the integrable systems do not obey strong ETH, weak ETH still holds, implying that all eigenstates look locally thermal, except for an exponentially small number of untypical ones \cite{Mori2016_09,Yoshizawa2018-120,Barthel2019_12}. Weak ETH applies very generally to many-body systems and only requires a sufficiently fast decay of correlations \cite{Biroli2010-105,Miao2019_05}. Two classes of interacting chains are investigated here with exact diagonalization: the Heisenberg XXZ model with or without next-nearest-neighbor interactions (Sec.~\ref{sec:XXZ}) and the transverse next-nearest-neighbor Ising model as well as its dual (Sec.~\ref{sec:tIsing}). For each case, we first assert the scaling properties by showing data collapse for subsystem entropies of thermodynamical ensembles. Then, we assess the applicability of ETH for entanglement entropies by computing entanglement for all eigenstates and comparing to the corresponding scaling function.

\section{Numerical simulations}
We use exact diagonalization \cite{Laflorencie2004-645,Sandvik2010-1297} for the models to obtain all $2^L$ eigenvalues and eigenvectors, where $L$ denotes the total system size. Employing periodic boundary conditions, the systems are translation invariant, reflection symmetric, and invariant under spin inversion. Hence, the Hamiltonians are block diagonal in a basis of semi-momentum states with crystal momentum $k$, reflection parity $p_\text{r}$, and spin-inversion parity $p_\text{s}$. The XXZ model also conserves the total $z$ magnetization $M$, which further reduces computation costs. Using full diagonalization in each block, we compute entanglement entropies $S(E,\ell)$ of all energy eigenstates [Eq.~\eqref{eq:S}]. For the test of ETH and the analysis of scaling properties, we compute thermal subsystem entropies
\begin{equation}\label{eq:Std}
	S_\td(\beta,\ell)=-\Tr \dm_{\td,\mc{A}} \ln{\dm_{\td,\mc{A}}}
\end{equation}
for the ensemble $\dm_\td=\frac{1}{Z}{{\rm e}}^{-\beta H}$ with inverse temperature $\beta$ and subsystem states $\dm_{\td,\mc{A}}=\Tr_\mc{B}\dm_\td$.

The largest system size in the simulations for the transverse Ising model is $L=18$, corresponding to 40 symmetry sectors. The largest two blocks have dimension 7314 each and quantum numbers $|k|=2\pi/3$, $p_\text{r}=\pm 1$, and $p_\text{s}=+1$.
The largest system size in the simulations for the XXZ model is $L=20$, corresponding to 436 symmetry sectors. The largest 36 blocks have dimension 8398 each and quantum numbers $|M|=1$, $|k|={\pi}/{10},\dotsc,{9\pi}/{10}$, $p_\text{r}=\pm 1$, and $p_\text{s}=\pm 1$.

To properly assess the scaling functions that describe the crossover from small $\ell$ and $E$ to large $\ell$ or $E$, one needs a sufficient number of low-energy states. Due to the computational restrictions on $L$, we hence restrict the analysis to critical points/phases of the spin models. Note, however, that eigenstate entanglement in gapped systems may also be described by scaling functions. As demonstrated in Ref.~\cite{Miao2019_05}, they feature additional parameters, e.g., the mass-temperature ratio.

\section{XXZ models}\label{sec:XXZ}
In this section, we consider the spin-1/2 XXZ model with (and without) competing antiferromagnetic next-nearest-neighbor interactions ($\alpha\geq 0$):
\begin{eqnarray}\label{eq:H-XXZ}
    H_{\text{XXZ}} = \sum_i \Big[\sigma^+_i \sigma^-_{i+1} + \sigma^-_i \sigma^+_{i+1} + \frac{\Delta}{2}\sigma^z_i \sigma^z_{i+1} \nonumber\\
    + \alpha \Big(\sigma^+_i \sigma^-_{i+2} + \sigma^-_i \sigma^+_{i+2} + \frac{|\Delta|}{2}\sigma^z_i \sigma^z_{i+2}\Big)\Big],
\end{eqnarray}
where $\sigma^{x,y,z}_i$ are the Pauli operators on site $i$, and $\sigma^{\pm}\equiv(\sigma^x\pm {{\rm i}} \sigma^y)/2$. Without next-nearest-neighbor terms ($\alpha = 0$), the model is Bethe-ansatz integrable \cite{Bethe1931,Korepin1993} but non-integrable otherwise. For $-1\leq\Delta\leq 1$, the spin-liquid phase of the nearest neighbor XXZ chain extends to nonzero $\alpha$. Increasing $\alpha$ further, 
a dimerized phase and a second spin-liquid phase follow \cite{Nomura1994-27,Somma2001-64}. For the numerical investigation, we choose the integrable point $(\Delta, \alpha) = (1/2,0)$, and the non-integrable point $(\Delta, \alpha) = (1/2,1/4)$ just below the transition to the dimerized phase which, for $\Delta=1/2$, occurs at $\alpha\approx 0.28$. The low-lying excitations in this spin-liquid phase are spinons and the long-range physics can be described by the sine-Gordon model, i.e., a conformal field theory (CFT) with central charge $c = 1$.

\subsection{Scaling of thermal subsystem entropy}
\begin{figure*}[t!]
\centering
	\includegraphics[width=0.93\columnwidth]{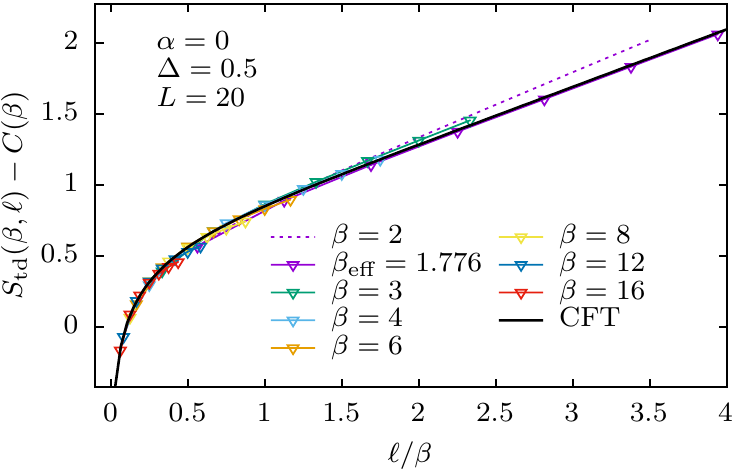}\hspace{7ex}
	\includegraphics[width=0.93\columnwidth]{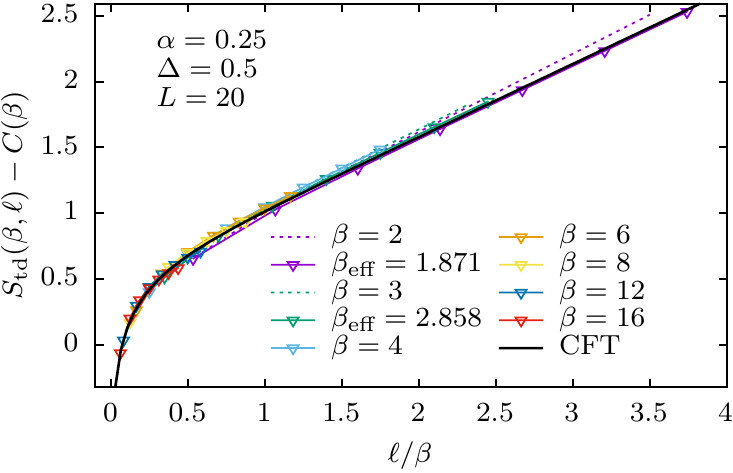}
	\caption{\label{fig:scaling_XXZ} \textbf{Crossover and scaling in XXZ chains.} The rescaled thermal subsystem entropies \eqref{eq:Std} for the integrable nearest-neighbor (left) and non-integrable next-nearest-neighbor (right) XXZ models \eqref{eq:H-XXZ} in the spin-liquid phase. Data is shown for total system size $L=20$, various inverse temperatures $\beta$, and subsystem sizes $\ell=1$ to $7$. After subtraction of the subleading term $C(\beta)$, the data collapses onto the crossover scaling function \eqref{eq:StdCFTscalingA} (black line), when plotted as a function of $\ell/\beta$.}
\end{figure*}
For the long-range physics of critical systems, there is just a single energy scale -- the temperature $\beta^{-1}$ -- and we hence expect a universal scaling function for the subsystem entropy \eqref{eq:Std} in the form $S_\td(\beta,\ell)\sim\Phi(\ell/\beta)$  \cite{Miao2019_05}. 
Independent of integrability, the low-energy excitations of the XXZ model in the spin-liquid phase are low-momentum spinons with a linear dispersion. Because of the resulting rotation and scale-invariance, the corresponding field theory is a 1+1d CFT \cite{Belavin1984-241,Francesco1997,Polchinski1988-383}. Within CFT, the thermal subsystem entropy can be computed with the replica trick and analytic continuation \cite{Korepin2004-92,Calabrese2004}. We obtain the scaling function \cite{Miao2019_05}
\begin{subequations}\label{eq:StdCFTscaling}
\begin{equation}\label{eq:StdCFTscalingA}
    \Phi^\cft(\ell/\beta)   =\frac{c}{3} \ln\left[\sinh\left(\frac{\pi}{v}\frac{\ell}{\beta}\right)\right],
\end{equation}
with the central charge $c$ and group velocity $v$ such that
\begin{equation}
	S^\cft_\td(\beta,\ell) = \Phi^\cft(\ell/\beta)+ C(\beta).
\end{equation}
\end{subequations}
Here, $C(\beta) = \frac{c}{3}\ln\left(\frac{\beta v}{\pi a}\right) + c^{\prime}$ is a subleading term with ultraviolet cutoff $1/a$ and  a nonuniversal constant $c'$.
For small subsystem size $\ell$ or small temperature $\beta^{-1}$, one recovers the celebrated log law $S(\ell)\sim\frac{c}{3}\ln \ell$ for groundstate entanglement \cite{Srednicki1993,Callan1994-333,Holzhey1994-424,Vidal2003-7,Jin2004-116,Calabrese2004,Zhou2005-12}. The crossover to a volume law occurs around $\ell\sim\ell_\text{c}:=\beta v/\pi$. For $\ell\gg\ell_\text{c}$, Eq.~\eqref{eq:StdCFTscaling} gives an extensive thermodynamic entropy
\begin{equation}\label{eq:StdExtensive}
	S^\cft_\td(\beta,\ell) \xrightarrow{\ell\gg\ell_\text{c}} \frac{c\pi\ell}{3v\beta}.
\end{equation}

To confirm this field-theoretic prediction, we numerically compute $S_\td(\beta,\ell)$ in the XXZ systems. The group velocity can be calculated from 
\begin{equation}\label{eq:velocity}
    v = \left[E_0(k_\text{gs}+\delta k) - E_0(k_\text{gs})\right]/\delta k,
\end{equation}
where $E_0(k)$ denotes the lowest energy in the symmetry sector with crystal momentum $k$, $k_\text{gs}$ is the groundstate momentum, and $\delta k=2\pi/L$ is the momentum spacing. We only consider even $L$. For $L=4n$, the groundstate sector has $k_\text{gs}=0$, and even reflection and spin-inversion parities. For $L=4n+2$, the groundstate has $k_\text{gs}=\pi$, and odd parities.

Figure~\ref{fig:scaling_XXZ} shows the results for both the integrable and non-integrable cases with various inverse temperatures $\beta$. The thermal subsystem entropies, plotted as functions of $\ell/\beta$ after subtraction of the subleading term $C(\beta)$, show indeed a collapse onto the crossover scaling function \eqref{eq:StdCFTscalingA}. The small deviations at the ends of the constant-$\beta$ lines are finite-size effects that become prominent for $\ell\approx L/2$. The curves for the highest considered temperatures ($\beta=2,3$) deviate somewhat towards larger $S_\td$. This is due to the fact that CFT, employed to derive Eq.~\eqref{eq:StdCFTscalingA}, assumes a perfectly linear dispersion for excitations. A linear dispersion up to arbitrarily high momenta can only occur in field-theoretical descriptions of condensed matter systems and results in ultraviolet divergences that need to be regularized. Lattice systems have a natural ultraviolet cutoff which results in a non-linear dispersion at higher energies. In particular, the excitation energy needs to have an extremum at the Brillouin-zone boundary.
As shown in Fig.~\ref{fig:scaling_XXZ}, this non-linearity can be compensated for almost entirely by defining an effective temperature through $\lim_{\ell\to\infty}S_\td(\beta,\ell)/\ell=:c\pi/[3v\beta_\eff(\beta)]$ and plotting with respect to $\ell/\beta_\eff(\beta)$ instead of $\ell/\beta$. As the CFT result is very precise for low temperatures where the non-linearity is irrelevant, Eq.~\eqref{eq:StdExtensive} implies that $\beta_\eff(\beta)\to\beta$ at low temperatures. For higher temperatures, $\beta_\eff$ has been defined such that replacing $\beta$ by $\beta_\eff$ in Eq.~\eqref{eq:StdExtensive} matches the exact thermodynamic entropy density.
In Fig.~\ref{fig:scaling_XXZ}, we employ this procedure only for the highest temperatures because $\beta_\eff$ is not needed for the lower temperatures, where its proper determination would also be complicated by the finite-size effects.

\subsection{Entanglement entropy and ETH}
\begin{figure*}[t!]
\centering
	{\small (a)}\includegraphics[width=0.9\columnwidth,valign=t]{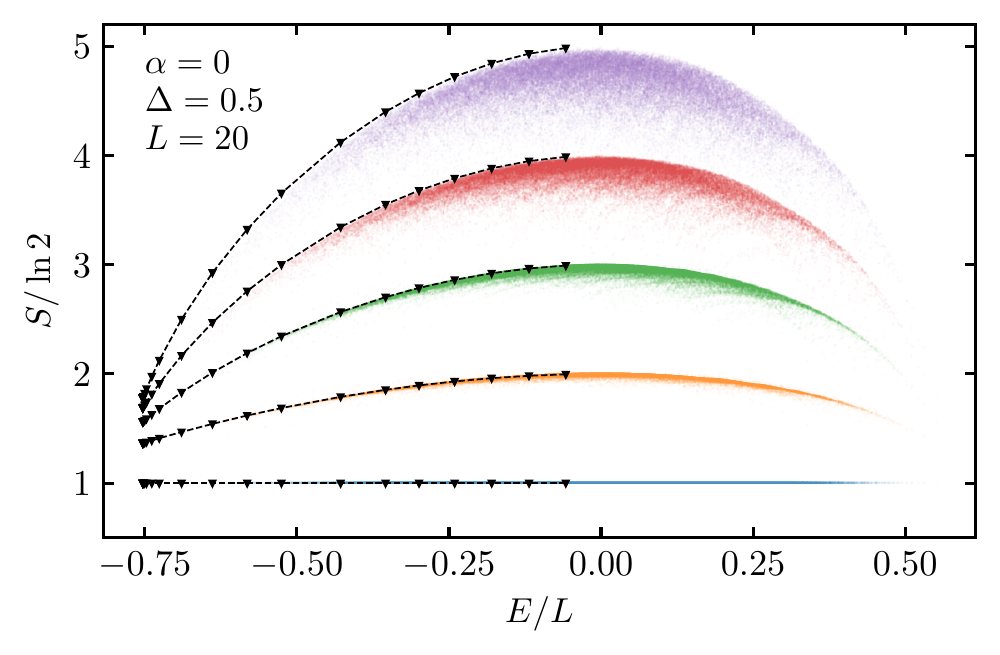}\hspace{4ex}
	{\small (b)}\includegraphics[width=0.9\columnwidth,valign=t]{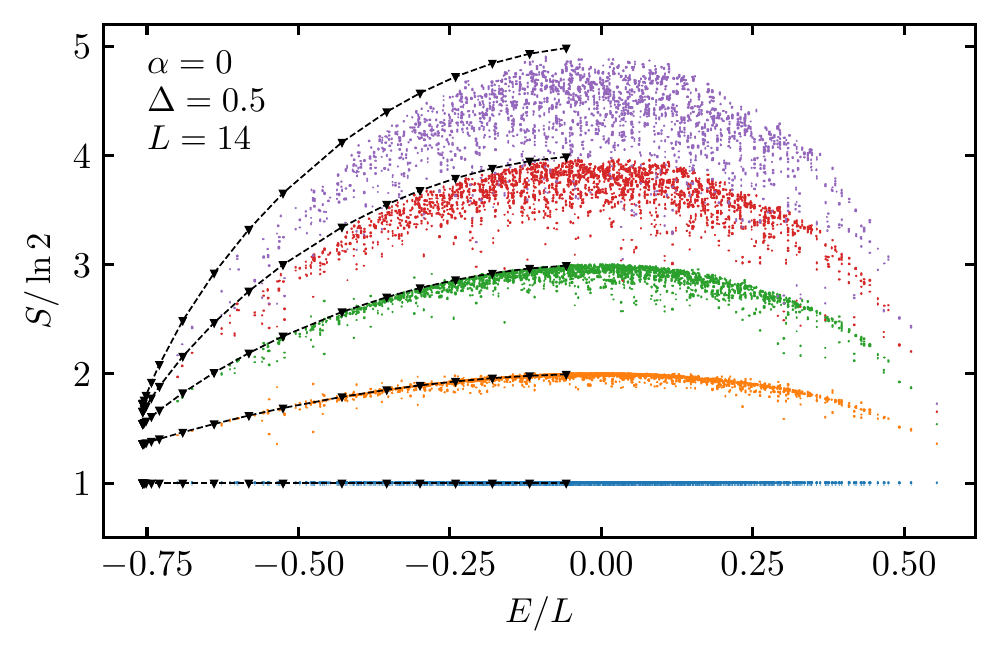}\\
	{\small (c)}\includegraphics[width=0.9\columnwidth,valign=t]{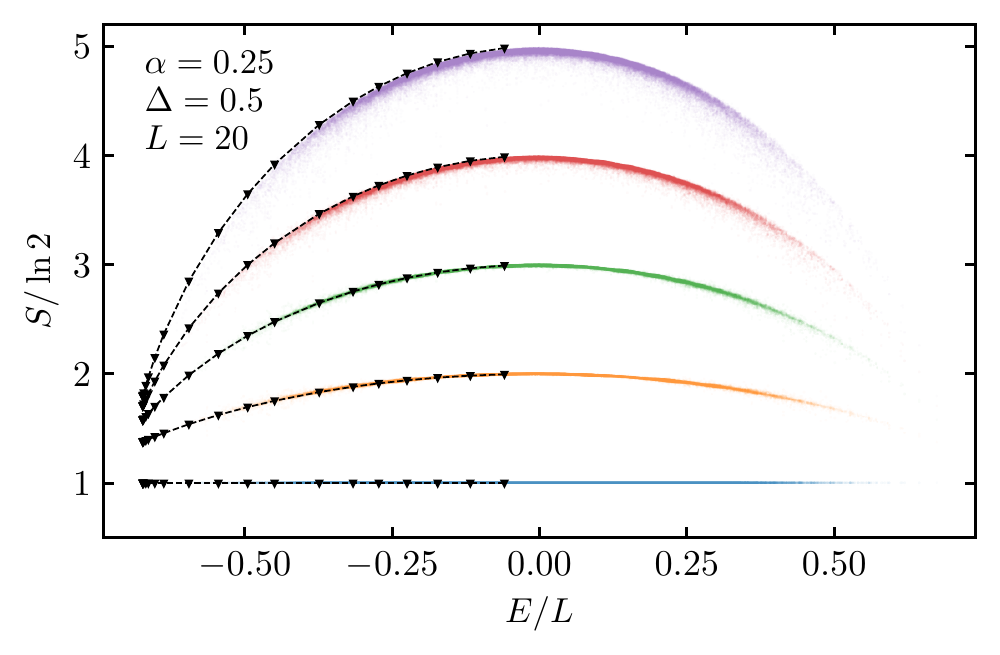}\hspace{4ex}
	{\small (d)}\includegraphics[width=0.9\columnwidth,valign=t]{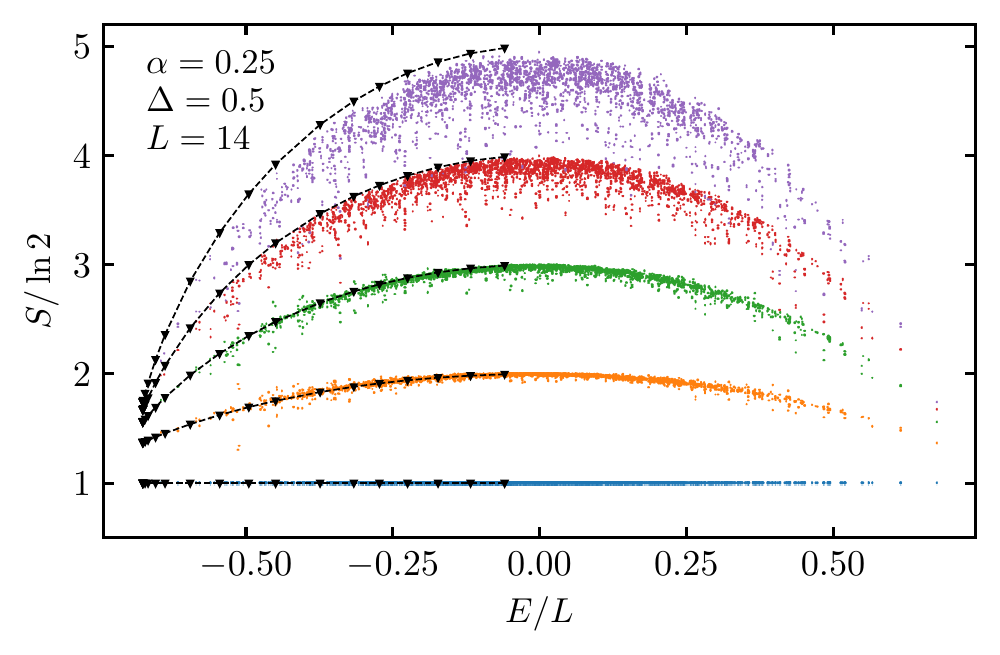}
	\caption{\label{fig:ETH_XXZ} \textbf{ETH for entanglement in XXZ chains.} The numerical simulations confirm that (weak or strong, respectively) ETH is applicable for the description of eigenstate entanglement entropies in both integrable (a,b) and non-integrable (c,d) XXZ models \eqref{eq:H-XXZ}. The dots give entanglement entropies $S(E,\ell)$ for all zero-magnetization energy eigenstates for subsystem sizes $\ell=1$ to $5$ (bottom to top). The larger the total system size $L$, the more they accumulate at the dashed lines which show the corresponding thermal subsystem entropies $S_\td(\beta(E),\ell)$.}
\end{figure*}
We have found that the thermal subsystem entropies $S_\td(\beta,\ell)$ obey the expected scaling behavior. Let us now probe whether ETH applies in the sense that the entanglement entropies $S(E,\ell)$ of (almost) all eigenstates converge to $S_\td(\beta,\ell)$ in the thermodynamic limit and are hence described by the same crossover scaling function \eqref{eq:StdCFTscaling}.

The XXZ model \eqref{eq:H-XXZ} conserves the total $z$ magnetization $M:=\sum_i\sigma^z_i/2$. Due to the spin-inversion symmetry, the considered thermodynamic ensemble $\dm_\td={\rm e}^{-\beta H}/Z$ in Eq.~\eqref{eq:Std} has a vanishing magnetization, $\Tr(\dm_\td M)=0$. Hence, we compare $S_\td$ to the entanglement entropies of all zero-magnetization eigenstates. Eigenstates with nonzero magnetization could, e.g., be compared with the canonical ensemble for the same $M$, or with the grand-canonical ensemble ${\rm e}^{-\beta(H-h M)}/Z$ with the right magnetization density fixed through the field $h$. In the thermodynamic limit, these ensembles are of course equivalent.

Figure~\ref{fig:ETH_XXZ} shows the numerical results for the critical XXZ model at the integrable point $(\Delta, \alpha) = (1/2,0)$ and the non-integrable point $(\Delta, \alpha) = (1/2,1/4)$. The entanglement entropies $S(E,\ell)$ of the energy eigenstates are plotted as semi-transparent dots for subsystem sizes $\ell=1$ to $5$. The thermal subsystem entropies $S_\td(\beta(E),\ell)$ are plotted as lines, where the temperature $\beta^{-1}(E)$ is chosen such that the energy expectation value for $\dm_\td$ matches $E$, i.e.,
\begin{equation}
	E=\frac{1}{Z}\Tr\big({\rm e}^{-\beta(E)H} H\big).
\end{equation}
For $\beta^{-1}\to\infty$, the energy approaches $E=0$ because $\Tr(\sigma^a_i \sigma^a_j)=0$ for any $i\neq j$ and $a=x,y,z$. Hence, all $S_\td(\beta(E),\ell)$ curves in Fig.~\ref{fig:ETH_XXZ} end at $E=0$. One would reach positive energies by considering negative temperatures or, equivalently, considering $-H$ instead of $H$. As this does not yield further insights, we refrain from doing so.

For fixed energy $E$, the distribution of $S(E,\ell)$ peaks at its upper edge that coincides well with $S_\td(\beta(E),\ell)$. The coincidence of the peak with the thermal subsystem entropy improves with increasing $L$ and decreasing $\ell$. The weight in the low-$S$ tail of the distribution decreases with increasing $L$. Not surprisingly, the low-$S$ tails are more prominent for the integrable systems. This is consistent with earlier results which established that standard deviations of observables decay exponentially in $L$ for non-integrable systems (strong ETH) \cite{Deutsch1991-43,Ikeda2011-84,Beugeling2014-89} and algebraically for integrable systems (weak ETH) \cite{Biroli2010-105,Ikeda2013-87,Alba2015-91,Barthel2019_12}. Similarly, large deviation bounds suggest that the ratio of untypical (athermal) eigenstates decreases double exponentially in $L$ for non-integrable systems and at least exponentially for integrable systems \cite{Mori2016_09,Yoshizawa2018-120,Barthel2019_12}. In Refs.~\cite{Miao2019_05,Barthel2019_12}, we simulated integrable non-interacting fermionic and bosonic models for which very large system sizes can be reached and found indeed $S(E,\ell)$ to be very sharply peaked at $S_\td(\beta(E),\ell)$.

\section{Next-nearest neighbor transverse Ising model and its dual}\label{sec:tIsing}
In this section, we study the 1d next-nearest-neighbor Ising model in a transverse field:
\begin{equation}\label{eq:H-tIsing}
    H_\tI = -\sum_{i}(J\sigma^z_i \sigma^z_{i+1} - \kappa\sigma^z_i \sigma^z_{i+2} + g\sigma^x_i).
\end{equation}
It is invariant under a $\pi$ rotation in the $xy$ plane ($\sigma^x\to-\sigma^x$, $\sigma^y\to-\sigma^y$) and $g\to -g$. Also,
the sign of the nearest-neighbor coupling $J$ is inessential as it can be altered by a $\pi$ rotation in the $yz$ plane ($\sigma^y\to-\sigma^y$, $\sigma^z\to-\sigma^z$) on every second site, which does not change the other two terms. Hence, we assume $g\geq 0$ and $J\geq 0$ without loss of generality.
\begin{figure}[t]
	\includegraphics[width=\columnwidth]{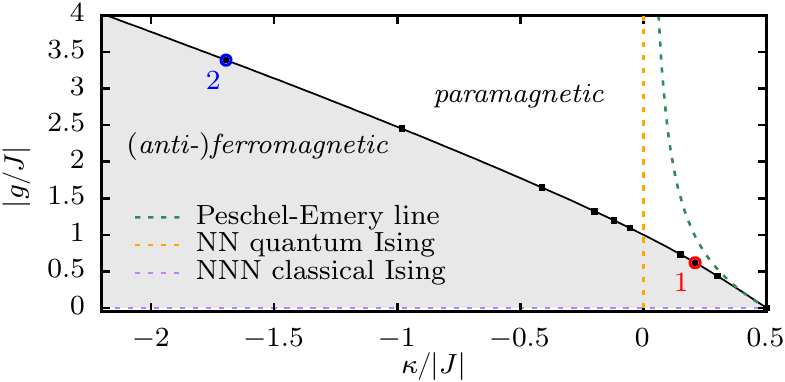}
	\caption{\label{fig:tIsing_phases} \textbf{Phase diagram of the quantum Ising chains.} For next-nearest-neighbor coupling $\kappa<J/2$, the model \eqref{eq:H-tIsing} features a ferromagnetic and a paramagnetic phase. Here, we determined the Ising transition line using ED and entanglement scaling. The model is integrable for $J=0$, for $\kappa=0$, and for $g=0$. It is frustration-free on the Peschel-Emery line \cite{Peschel1981-43}. For the investigation of eigenstate entanglement, we choose the indicated critical points $(\kappa_1/J,g_1/J)=(0.21,0.62)$ and $(\kappa_2/J,g_2/J)=(-1.695,3.39)$.}
\end{figure}

In the framework of the quantum-classical correspondence \cite{Sachdev2011}, the model \eqref{eq:H-tIsing} is closely related to the classical axial next-nearest-neighbor Ising (ANNNI) model in 2d, and hence often referred to as the quantum ANNNI model. For $\kappa<J/2$, there exists a second-order Ising-type quantum phase transition between a ferromagnetic phase at small $g$ and a paramagnetic phase at large $g$ as shown in Fig.~\ref{fig:tIsing_phases}. For $\kappa>J/2$ there is the so-called antiphase at small $g$, followed by the critical floating phase, before transitioning to the paramagnetic phase \cite{Fisher1980-44,Rujan1981-24,Peschel1981-43,Bak1982-45,Selke1988-170,Allen2001-34,Beccaria2006-73,Beccaria2007-76,Sela2011-84,Dutta2015}. All four phases meet at the multicritical point $(\kappa,g)=(J/2,0)$. The model is non-integrable except for the three lines $J=0$, $g=0$, and $\kappa=0$. And it is frustration-free on the Peschel-Emery one-dimensional line $g=J^2/(4\kappa)-\kappa$ \cite{Peschel1981-43,Beccaria2006-73}.
Here, we set $J=1$ and choose the two points  $(\kappa_1,g_1)=(0.21,0.62)$ and $(\kappa_2,g_2)=(-1.695,3.39)$ on the transition line between the ferromagnetic and paramagnetic phases. Both are described by an Ising CFT with central charge  $c = 1/2$.

Ising systems are also often discussed in terms of dual bond variables \cite{Fradkin2013}
\begin{equation}\label{eq:bondVariables}
	 \tau^x_i:=\sigma^z_i\sigma^z_{i+1},\quad	 
	 \tau^z_i:=\prod_{j\leq i}\sigma^x_j.
\end{equation}
Together with $\tau^y_i:={\rm i}\tau^x_i\tau^z_i$, they obey the Pauli matrix algebra. In terms of these operators, the model \eqref{eq:H-tIsing} reads
\begin{equation}\label{eq:H-tIsingD}
	H'_\tI = -\sum_i (J \tau^x_i -\kappa \tau^x_i \tau^x_{i+1} + g\tau^z_i \tau^z_{i+1}).
\end{equation}
This dual Hamiltonian has only nearest-neighbor interactions and comprises the integrable transverse Ising and XY models. In this form, the model has, for example, been discussed in Refs.~\cite{Peschel1981-43,Hassler2012-14,Cole2017-95,Mahyaeh2020-101}. While the structure of the phase diagram remains the same, the physical interpretation of the various phases changes. In particular, the paramagnetic phase becomes a topological phase with two ground states distinguished by parity. With the Jordan-Wigner transformation from $\{\tau^{x,y,z}_i\}$ to fermionic operators \cite{Jordan1928}, $H'_\tI$ becomes the Kitaev-Hubbard chain \cite{Mahyaeh2020-101}.
\begin{figure}[t]
	\includegraphics[width=\columnwidth]{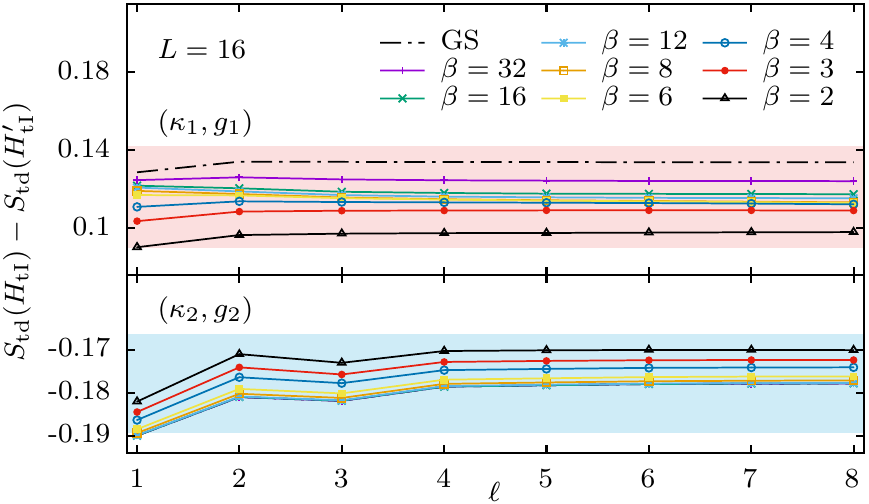}
	\caption{\label{fig:dual_diff} \textbf{Difference in subsystem entropies for the Ising model and its dual.} For the same system parameters, subsystem entropies \eqref{eq:Std} of the next-nearest-neighbor Ising model \eqref{eq:H-tIsing} and its dual \eqref{eq:H-tIsingD} are very similar and converge to a constant for large subsystem size $\ell$.}
\end{figure}

\subsection{Scaling of thermal subsystem entropy}
\begin{figure*}[t]
\centering
        \includegraphics[width=0.93\columnwidth]{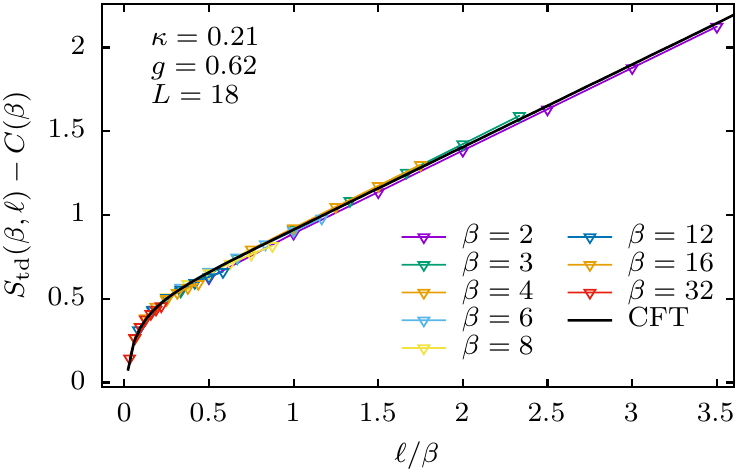}\hspace{7ex}
	\includegraphics[width=0.93\columnwidth]{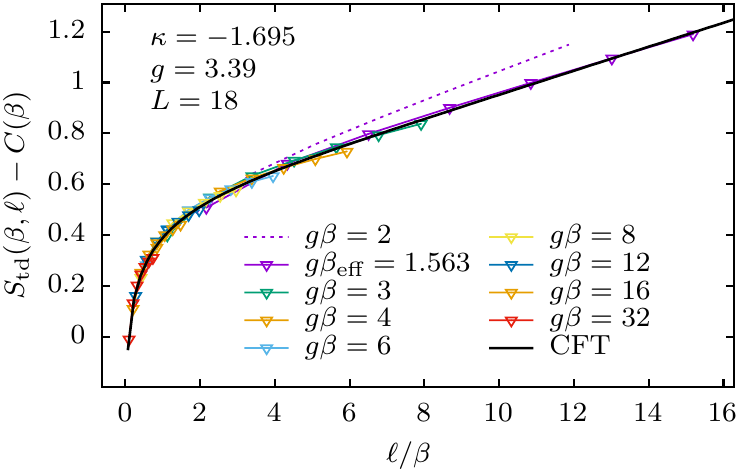}
	\caption{\label{fig:scaling_Ising} \textbf{Crossover and scaling in quantum Ising chains and their duals.} The rescaled thermal subsystem entropies \eqref{eq:Std} for the next-nearest-neighbor Ising model $H_\tI$ (left) and its dual $H'_\tI$ (right) at two critical points. Data is shown for $L=18$, and subsystem sizes $\ell=1$ to $7$. With proper rescaling, the data collapses onto the crossover scaling function \eqref{eq:StdCFTscalingA} (black line).}
\end{figure*}
The mapping \eqref{eq:bondVariables} from $H_\tI$ to the dual $H'_\tI$ is actually an exact unitary transformation if one applies open boundary conditions in the former and suitable fixed boundary conditions in the latter. An up-spin in the original representation maps to a domain wall. Except for unitary string operators, the mapping is local, and the entanglement properties of both models are hence closely related. For the simulations, we employ periodic boundary conditions. While they do not affect the local physics and the structure of the phase diagram, they affect entanglement and thermal subsystem entropies due to boundary contributions. Figure~\ref{fig:dual_diff} shows that the differences of subsystem entropies of the model \eqref{eq:H-tIsing} and its dual \eqref{eq:H-tIsingD} are in fact almost independent of the subsystem size and have only a small temperature dependence. In the following, we study the point $(\kappa_1,g_1)$ in the next-nearest-neighbor model $H_\tI$ and the second point, $(\kappa_2,g_2)$, in the dual model $H'_\tI$ to make sure that ETH and the general scaling arguments from the introduction and Ref.~\cite{Miao2019_05} work in both representations.
\begin{figure*}[t!]
	{\small (a)}\includegraphics[width=0.89\columnwidth,valign=t]{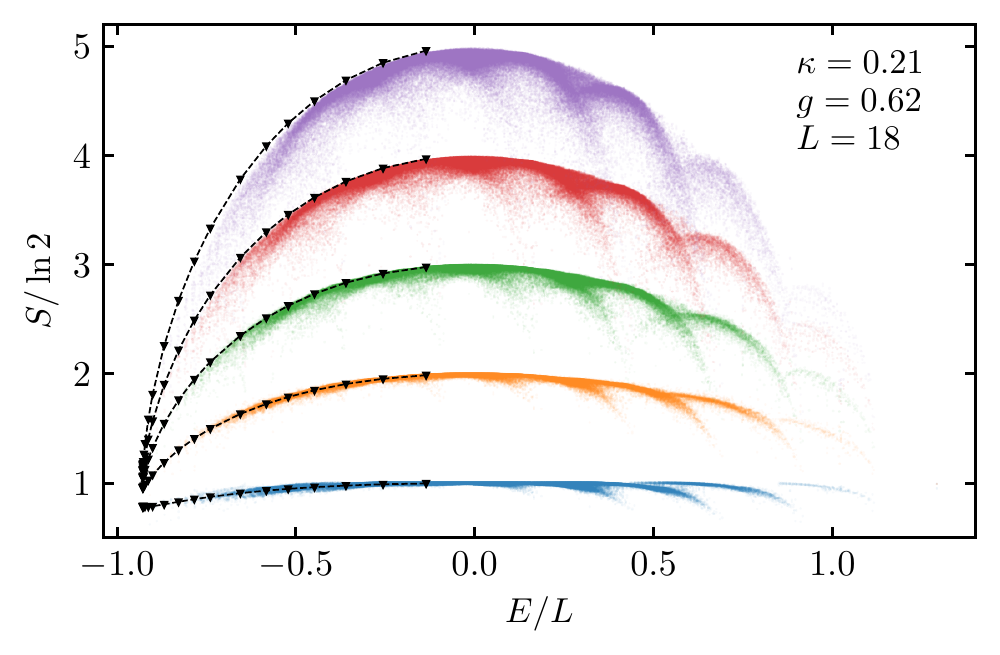}\hspace{4ex}
	{\small (b)}\includegraphics[width=0.89\columnwidth,valign=t]{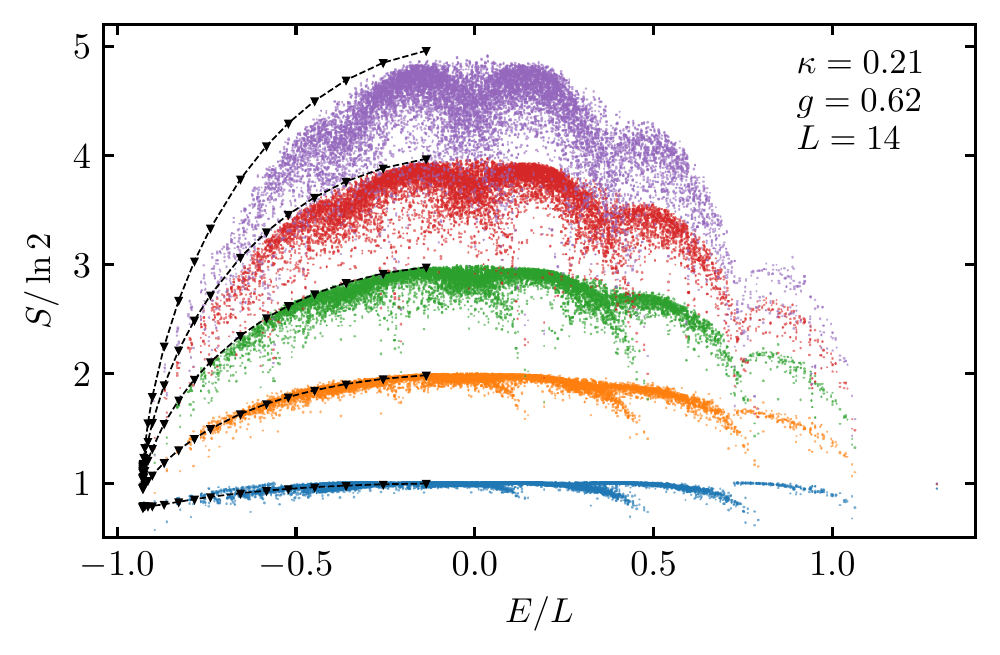}\\
	{\small (c)}\includegraphics[width=0.89\columnwidth,valign=t]{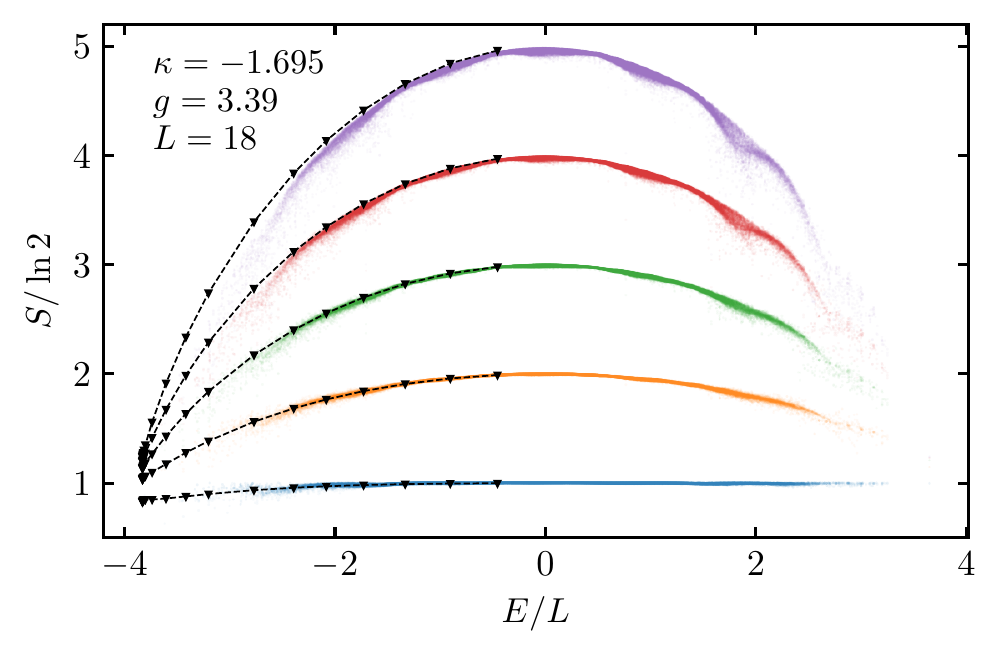}\hspace{4ex}
	{\small (d)}\includegraphics[width=0.89\columnwidth,valign=t]{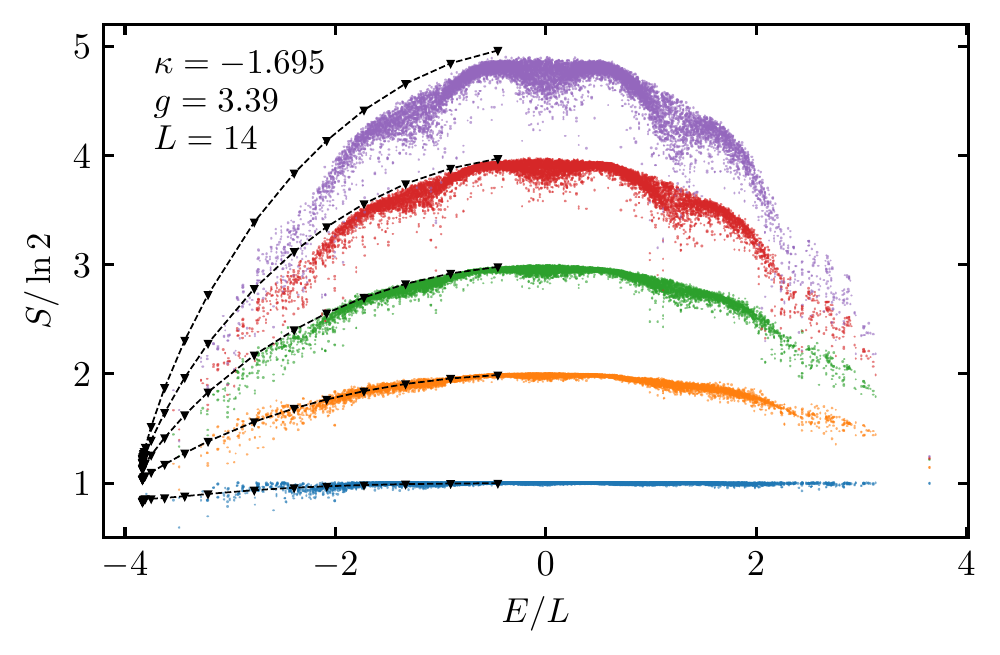}
	\caption{\label{fig:ETH_Ising} \textbf{ETH for entanglement in quantum Ising chains and their duals.} The numerical simulations confirm that ETH is applicable to eigenstate entanglement entropies in the non-integrable next-nearest-neighbor transverse Ising model $H_\tI$ (panels \emph{a} and \emph{b}) and its dual $H'_\tI$ (panels \emph{c} and \emph{d}). The dots show entanglement entropies $S(E,\ell)$ of all energy eigenstates for subsystem sizes $\ell=1$ to $5$ (bottom to top), at the two critical points indicated in Fig.~\ref{fig:tIsing_phases}. The larger the total system size $L$, the more they accumulate at the dashed lines which show the corresponding thermal subsystem entropies $S_\td(\beta(E),\ell)$.}
\end{figure*}
\begin{figure*}[t!]
	{\small (a)}\includegraphics[width=0.92\columnwidth,valign=t]{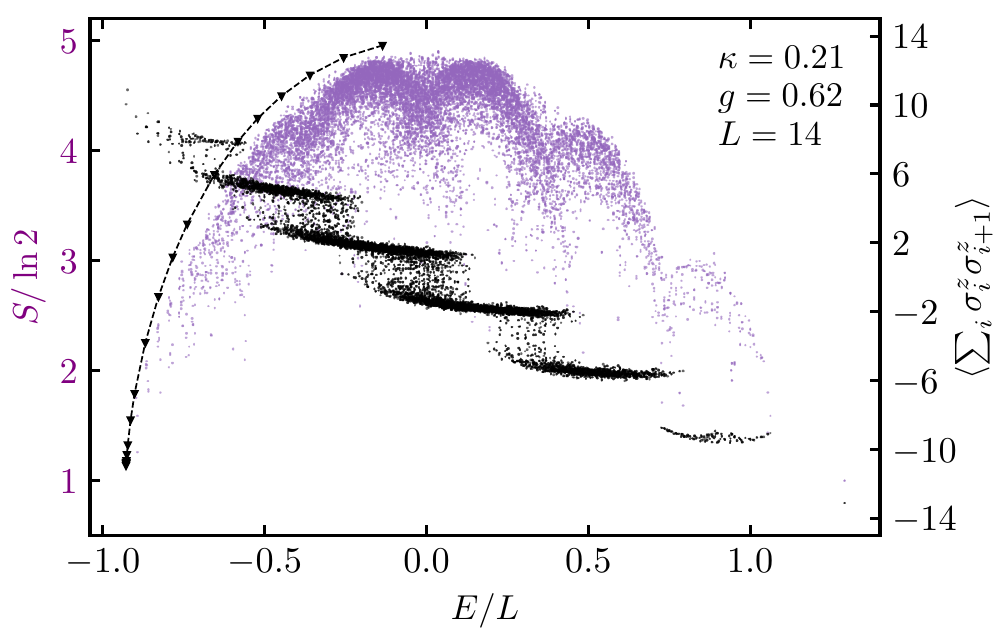}\hspace{4ex}
	{\small (b)}\includegraphics[width=0.92\columnwidth,valign=t]{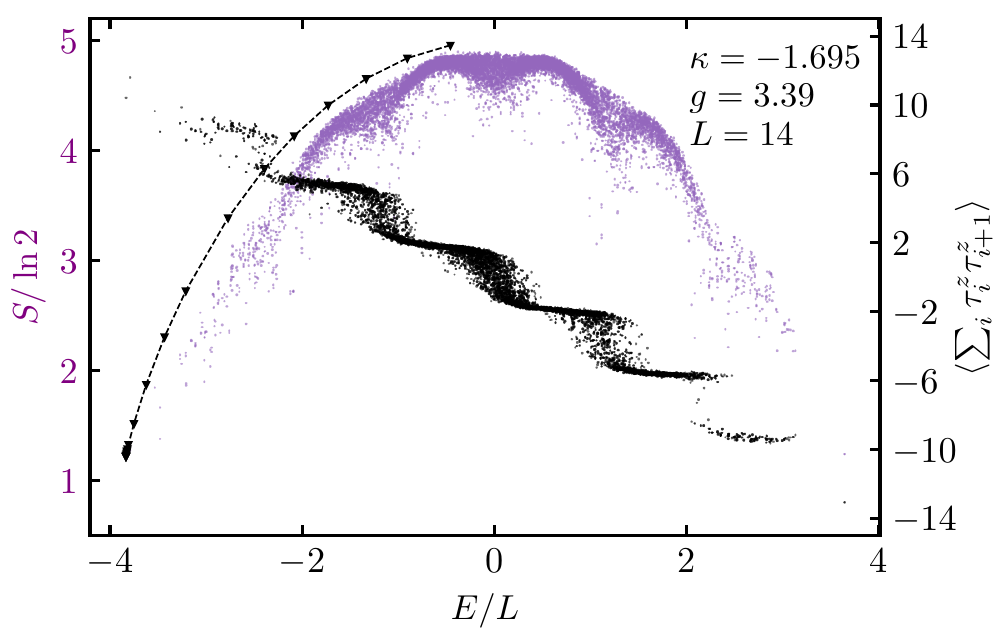}
	\caption{\label{fig:ETH_Ising-arcs} \textbf{Explanation of entanglement arcs in the quantum Ising chains.} The eigenstate entanglement shown in Fig.~\ref{fig:ETH_Ising-arcs} displays arc-like substructures that emerge below the curve for the thermal subsystem entropy. They are more pronounced for smaller system sizes $L$. In the text we explain these features by approximately conserved quantities. For total system size $L=14$ and subsystem size $\ell=5$, the plots show eigenstate entanglement entropies $S(E,\ell)$ in the next-nearest-neighbor transverse Ising model $H_\tI$ (panel \emph{a}) and its dual $H'_\tI$ (panel \emph{b}) at the two critical points indicated in Fig.~\ref{fig:tIsing_phases}. They are compared to expectation values of the number of $\sigma^z$ and $\tau^z$ domain walls, quantified by $\sum_i \sigma^z_i\sigma^z_{i+1}$ and $\sum_i \tau^z_i\tau^z_{i+1}$, respectively.}
\end{figure*}

The group velocity $v$ for the low-energy excitations is again determined according to Eq.~\eqref{eq:velocity} in the groundstate sector which, here, has momentum $k_\text{gs}=0$ and even reflection and spin-inversion parities.
Figure~\ref{fig:scaling_Ising} shows the thermal subsystem entropies for various inverse temperatures $\beta$. When subtracting the subleading constant $C(\beta)$ and plotting as a function of $\ell/\beta$, the data collapses very well onto the crossover scaling function \eqref{eq:StdCFTscaling}. Similar to the observations in the XXZ model, for the point $(\kappa_2,g_2)$ at the highest temperature ($g\beta=2$), one sees some deviation from the scaling function due to the nonlinearity of the dispersion at higher energies. This effect can again be compensated for by employing the effective temperature $\beta^{-1}_\eff$.

\subsection{Entanglement entropy and ETH}
Figure~\ref{fig:ETH_Ising} is a scatter plot of eigenstate entanglement entropies $S(E,\ell)$ for the next-nearest-neighbor transverse Ising model $H_\tI$ and its dual $H'_\tI$ for the two points on the transition line between the ferromagnetic and paramagnetic phases indicated in Fig.~\ref{fig:tIsing_phases}. For comparison, the corresponding thermal subsystem entropies $S_\td(\beta(E),\ell)$ are plotted as lines.
For $H'_\tI$ at point $(\kappa_2,g_2)$ [panels \emph{c} and \emph{d}], the distributions of eigenstate entanglement are very sharp in the bulk of the spectrum and coincide well with $S_\td(\beta(E),\ell)$. At low energies, the distributions get broader but tighten with increasing $L$ and decreasing $\ell$.
For $H_\tI$ at $(\kappa_1,g_1)$ [panels \emph{a} and \emph{b}], the eigenstate entanglement distributions peak at their upper edges which still coincide well with $S_\td(\beta(E),\ell)$, and the weight in the low-$S$ tails decreases with increasing $L$. But the distributions are generally broader than for $(\kappa_2,g_2)$. This should be due to the fact that the point $(\kappa_1,g_1)$ is relatively close to the two integrable lines at $g=0$ and $\kappa=0$, and to the frustration-free Peschel-Emery line; see Fig.~\ref{fig:tIsing_phases}. So strong ETH should apply in the thermodynamic limit, but, at small $L$, deviations from the thermal subsystem entropies can be similarly big as in integrable systems, which only obey weak ETH.

All four plots display arc-like substructures in the eigenstate entanglement, which are more pronounced for small $L$. Had we not chosen to only plot zero-magnetization data in Fig.~\ref{fig:ETH_XXZ}, similar structures would occur for the XXZ model with one arc for every magnetization sector. In the thermodynamic limit, the magnetization density of almost all $H_{\text{XXZ}}$ eigenstates is in a small interval $[-\epsilon,\epsilon]$ around zero (concentration of measure) and, hence, the arc-like structures for the XXZ model disappear for large $L$. The transverse Ising model $H_\tI$ and its dual $H'_\tI$ do not have symmetries except for translation, spatial reflection, and spin inversion, which do not explain the arc-like structures in Fig.~\ref{fig:ETH_Ising}. However, for $(\kappa,g)=(0.21,0.62)$, the transverse Ising model \eqref{eq:H-tIsing} is relatively close to the line $g=0$, where the total $z$ magnetization is conserved and all eigenstates can be chosen as tensor products of $\sigma_i^z$ eigenstates. To first order in $g$, one needs to superimpose each such state $|\psi\ket$ with the spin-flipped state $\bigotimes_i\sigma_i^x|\psi\ket$ to obtain the perturbed eigenstates. While the resulting states have zero magnetization, they have a fixed number of $\sigma^z$ domain walls. The eigenstate expectation values of $\sum_i\sigma^z_i\sigma^z_{i+1}$ shown in Fig.~\ref{fig:ETH_Ising-arcs}a establish that the number of $\sigma^z$ domain walls is indeed approximately conserved and corresponds exactly to the entanglement arcs. Similar to the case of the XXZ model, the arc features will vanish with increasing $L$ due to the concentration of measure. The situation of the dual model \eqref{eq:H-tIsingD} is analogous. For $(\kappa,g)=(-1.695,3.39)$, the term $\tau^z_i\tau^z_{i+1}$ dominates and $\sum_i\tau^z_i\tau^z_{i+1}$ is approximately conserved, explaining the arc-like structures in Fig.~\ref{fig:ETH_Ising-arcs}b.

\section{Conclusion}
In summary, we have numerically confirmed eigenstate entanglement scaling for critical interacting 1d systems. For (almost) all energy eigenstates, the crossover from the groundstate scaling at low energies $E$ and small subsystem sizes $\ell$ to a volume law at large $E$ or $\ell$ is captured by a universal scaling function derived from CFT. Both integrable and non-integrable models were investigated using exact diagonalization. Although the accessible system size is limited for these interacting systems, the thermal subsystem entropies match the predicted scaling function very well, and the results establish the applicability of ETH for the entanglement entropies. We also shortly addressed differences in subsystem entropies for the next-nearest-neighbor transverse Ising model and its dual. Due to boundary effects, they basically differ by a constant with small temperature dependence. The results here complement numerical investigations of eigenstate entanglement for large non-interacting fermionic and bosonic systems in Refs.~\cite{Miao2019_05,Barthel2019_12}.

\begin{acknowledgments}
We gratefully acknowledge helpful discussions with Vincenzo Alba, Pasquale Calabrese, Anatoly Dymarsky, Isreal Klich, Jianfeng Lu, Shang-Qiang Ning, Marcos Rigol, and Yikang Zhang, as well as support through the US Department of Energy grant DE-SC0019449.
\end{acknowledgments}


\begin{thebibliography}{10}

\bibitem{Amico2008-80}
L. Amico, R. Fazio, A. Osterloh, and V. Vedral, {\em Entanglement in many-body
  systems}, \href{https://doi.org/10.1103/RevModPhys.80.517} {Rev. Mod. Phys.
  {\bf 80},  517  (2008)}.

\bibitem{Horodecki2009-81}
R. Horodecki, P. Horodecki, M. Horodecki, and K. Horodecki, {\em Quantum
  entanglement}, \href{https://doi.org/10.1103/RevModPhys.81.865} {Rev. Mod.
  Phys. {\bf 81},  865  (2009)}.

\bibitem{Laflorencie2016-646}
N. Laflorencie, {\em Quantum entanglement in condensed matter systems},
  \href{https://doi.org/10.1016/j.physrep.2016.06.008} {Phys. Rep. {\bf 646},
  1   (2016)}.

\bibitem{Eisert2008}
J. Eisert, M. Cramer, and M.~B. Plenio, {\em Colloquium: Area laws for the
  entanglement entropy}, \href{https://doi.org/10.1103/RevModPhys.82.277} {Rev.
  Mod. Phys. {\bf 82},  277  (2010)}.

\bibitem{Latorre2009}
J.~I. Latorre and A. Riera, {\em A short review on entanglement in quantum spin
  systems}, \href{https://doi.org/10.1088/1751-8113/42/50/504002} {J. Phys. A:
  Math. Theor. {\bf 42},  504002  (2009)}.

\bibitem{Srednicki1993}
M. Srednicki, {\em Entropy and area},
  \href{https://doi.org/10.1103/PhysRevLett.71.666} {Phys. Rev. Lett. {\bf 71},
   666  (1993)}.

\bibitem{Callan1994-333}
C. Callan and F. Wilczek, {\em On geometric entropy},
  \href{https://doi.org/doi:10.1016/0370-2693(94)91007-3} {Phys. Lett. B {\bf
  333},  55  (1994)}.

\bibitem{Latorre2004}
J.~I. Latorre, E. Rico, and G. Vidal, {\em Ground state entanglement in quantum
  spin chains}, {Quantum Info. Comput. {\bf 4},  48  (2004)}.

\bibitem{Plenio2005}
M.~B. Plenio, J. Eisert, J. Drei\ss{}ig, and M. Cramer, {\em Entropy,
  entanglement, and area: analytical results for harmonic lattice systems},
  \href{https://doi.org/10.1103/PhysRevLett.94.060503} {Phys. Rev. Lett. {\bf
  94},  060503  (2005)}.

\bibitem{Cramer2006-73}
M. Cramer, J. Eisert, M.~B. Plenio, and J. Drei\ss{}ig, {\em Entanglement-area
  law for general bosonic harmonic lattice systems},
  \href{https://doi.org/10.1103/PhysRevA.73.012309} {Phys. Rev. A {\bf 73},
  012309  (2006)}.

\bibitem{Hastings2007-76}
M.~B. Hastings, {\em Entropy and entanglement in quantum ground states},
  \href{https://doi.org/10.1103/PhysRevB.76.035114} {Phys. Rev. B {\bf 76},
  035114  (2007)}.

\bibitem{Brandao2013-9}
F.~G. S.~L. Brand{\~a}o and M. Horodecki, {\em An area law for entanglement
  from exponential decay of correlations},
  \href{https://doi.org/10.1038/nphys2747} {Nat. Phys. {\bf 9},  721  (2013)}.

\bibitem{Cho2018-8}
J. Cho, {\em Realistic area-law bound on entanglement from exponentially
  decaying correlations}, \href{https://doi.org/10.1103/PhysRevX.8.031009}
  {Phys. Rev. X {\bf 8},  031009  (2018)}.

\bibitem{Holzhey1994-424}
C. Holzhey, F. Larsen, and F. Wilczek, {\em Geometric and renormalized entropy
  in conformal field theory},
  \href{https://doi.org/10.1016/0550-3213(94)90402-2} {Nucl. Phys. B {\bf 424},
   443  (1994)}.

\bibitem{Vidal2003-7}
G. Vidal, J.~I. Latorre, E. Rico, and A. Kitaev, {\em Entanglement in quantum
  critical phenomena}, \href{https://doi.org/10.1103/PhysRevLett.90.227902}
  {Phys. Rev. Lett. {\bf 90},  227902  (2003)}.

\bibitem{Jin2004-116}
B.~Q. Jin and V.~E. Korepin, {\em Quantum spin chain, Toeplitz determinants and
  Fisher-Hartwig conjecture},
  \href{https://doi.org/10.1023/B:JOSS.0000037230.37166.42} {J. Stat. Phys.
  {\bf 116},  79  (2004)}.

\bibitem{Calabrese2004}
P. Calabrese and J.~L. Cardy, {\em Entanglement entropy and quantum field
  theory}, \href{https://doi.org/10.1088/1742-5468/2004/06/P06002} {J. Stat.
  Mech.  P06002  (2004)}.

\bibitem{Zhou2005-12}
H.-Q. Zhou, T. Barthel, J.~O. Fj{\ae}restad, and U. Schollw\"ock, {\em
  Entanglement and boundary critical phenomena},
  \href{https://doi.org/10.1103/PhysRevA.74.050305} {Phys. Rev. A {\bf 74},
  050305(R)  (2006)}.

\bibitem{Wolf2005}
M.~M. Wolf, {\em Violation of the entropic area law for fermions},
  \href{https://doi.org/10.1103/PhysRevLett.96.010404} {Phys. Rev. Lett. {\bf
  96},  010404  (2006)}.

\bibitem{Gioev2005}
D. Gioev and I. Klich, {\em Entanglement entropy of fermions in any dimension
  and the {W}idom conjecture},
  \href{https://doi.org/10.1103/PhysRevLett.96.100503} {Phys. Rev. Lett. {\bf
  96},  100503  (2006)}.

\bibitem{Barthel2006-74}
T. Barthel, M.-C. Chung, and U. Schollw\"ock, {\em Entanglement scaling in
  critical two-dimensional fermionic and bosonic systems},
  \href{https://doi.org/10.1103/PhysRevA.74.022329} {Phys. Rev. A {\bf 74},
  022329  (2006)}.

\bibitem{Li2006}
W. Li, L. Ding, R. Yu, T. Roscilde, and S. Haas, {\em Scaling behavior of
  entanglement in two- and three-dimensional free-fermion systems},
  \href{https://doi.org/10.1103/PhysRevB.74.073103} {Phys. Rev. B {\bf 74},
  073103  (2006)}.

\bibitem{Lai2013-111}
H.-H. Lai, K. Yang, and N.~E. Bonesteel, {\em Violation of the entanglement
  area law in bosonic systems with Bose surfaces: Possible application to Bose
  metals}, \href{https://doi.org/10.1103/PhysRevLett.111.210402} {Phys. Rev.
  Lett. {\bf 111},  210402  (2013)}.

\bibitem{Murciano2020-8}
S. Murciano, P. Ruggiero, and P. Calabrese, {\em Symmetry resolved entanglement
  in two-dimensional systems via dimensional reduction},
  \href{https://doi.org/10.1088/1742-5468/aba1e5} {J. Stat. Mech.  083102
  (2020)}.

\bibitem{Popescu2006-2}
S. Popescu, A.~J. Short, and A. Winter, {\em Entanglement and the foundations
  of statistical mechanics}, \href{https://doi.org/doi:10.1038/nphys444} {Nat.
  Phys. {\bf 2},  754  (2006)}.

\bibitem{Goldstein2006-96}
S. Goldstein, J.~L. Lebowitz, R. Tumulka, and N. Zangh\`{\i}, {\em Canonical
  typicality}, \href{https://doi.org/10.1103/PhysRevLett.96.050403} {Phys. Rev.
  Lett. {\bf 96},  050403  (2006)}.

\bibitem{Gemmer2004}
J. Gemmer, M. Michel, and G. Mahler, {\em Quantum Thermodynamics}, Vol.~657 of
  {\em Lecture Notes in Physics} (Springer, Berlin, Heidelberg, 2004).

\bibitem{Alba2009-10}
V. Alba, M. Fagotti, and P. Calabrese, {\em Entanglement entropy of excited
  states}, \href{https://doi.org/10.1088/1742-5468/2009/10/p10020} {J. Stat.
  Mech.  P10020  (2009)}.

\bibitem{Ares2014-47}
F. Ares, J.~G. Esteve, F. Falceto, and E. S{\'{a}}nchez-Burillo, {\em Excited
  state entanglement in homogeneous fermionic chains},
  \href{https://doi.org/10.1088/1751-8113/47/24/245301} {J. Phys. A: Math.
  Theor. {\bf 47},  245301  (2014)}.

\bibitem{Storms2014-89}
M. Storms and R.~R.~P. Singh, {\em Entanglement in ground and excited states of
  gapped free-fermion systems and their relationship with Fermi surface and
  thermodynamic equilibrium properties},
  \href{https://doi.org/10.1103/PhysRevE.89.012125} {Phys. Rev. E {\bf 89},
  012125  (2014)}.

\bibitem{Moelter2014-10}
J. M\"olter, T. Barthel, U. Schollw\"ock, and V. Alba, {\em Bound states and
  entanglement in the excited states of quantum spin chains},
  \href{https://doi.org/10.1088/1742-5468/2014/10/P10029} {J. Stat. Mech.
  P10029  (2014)}.

\bibitem{Keating2015-338}
J.~P. Keating, N. Linden, and H.~J. Wells, {\em Spectra and eigenstates of spin
  chain Hamiltonians}, \href{https://doi.org/10.1007/s00220-015-2366-0}
  {Commun. Math. Phys. {\bf 338},  81  (2015)}.

\bibitem{Vdimar2017-119}
L. Vidmar, L. Hackl, E. Bianchi, and M. Rigol, {\em Entanglement entropy of
  eigenstates of quadratic fermionic Hamiltonians},
  \href{https://doi.org/10.1103/PhysRevLett.119.020601} {Phys. Rev. Lett. {\bf
  119},  020601  (2017)}.

\bibitem{Vidmar2017-119b}
L. Vidmar and M. Rigol, {\em Entanglement entropy of eigenstates of quantum
  chaotic Hamiltonians}, \href{https://doi.org/10.1103/PhysRevLett.119.220603}
  {Phys. Rev. Lett. {\bf 119},  220603  (2017)}.

\bibitem{Vidmar2018-121}
L. Vidmar, L. Hackl, E. Bianchi, and M. Rigol, {\em Volume law and quantum
  criticality in the entanglement entropy of excited eigenstates of the quantum
  Ising model}, \href{https://doi.org/10.1103/PhysRevLett.121.220602} {Phys.
  Rev. Lett. {\bf 121},  220602  (2018)}.

\bibitem{Lu2019-99}
T.-C. Lu and T. Grover, {\em Renyi entropy of chaotic eigenstates},
  \href{https://doi.org/10.1103/PhysRevE.99.032111} {Phys. Rev. E {\bf 99},
  032111  (2019)}.

\bibitem{Huang2019-938}
Y. Huang, {\em Universal eigenstate entanglement of chaotic local
  Hamiltonians}, \href{https://doi.org/10.1016/j.nuclphysb.2018.09.013} {Nucl.
  Phys. B {\bf 938},  594   (2019)}.

\bibitem{LeBlond2019-100}
T. LeBlond, K. Mallayya, L. Vidmar, and M. Rigol, {\em Entanglement and matrix
  elements of observables in interacting integrable systems},
  \href{https://doi.org/10.1103/PhysRevE.100.062134} {Phys. Rev. E {\bf 100},
  062134  (2019)}.

\bibitem{Lydzba2020_06}
P. \L{}yd\.{z}ba, M. Rigol, and L. Vidmar, {\em Eigenstate entanglement entropy
  in random quadratic Hamiltonians},
  \href{https://doi.org/10.1103/PhysRevLett.125.180604} {Phys. Rev. Lett. {\bf
  125},  180604  (2020)}.

\bibitem{Miao2019_05}
Q. Miao and T. Barthel, {\em Eigenstate entanglement: Crossover from the ground
  state to volume laws}, \href{https://doi.org/10.1103/PhysRevLett.127.040603}
  {Phys. Rev. Lett. {\bf 127},  040603  (2021)}.

\bibitem{Deutsch1991-43}
J.~M. Deutsch, {\em Quantum statistical mechanics in a closed system},
  \href{https://doi.org/10.1103/PhysRevA.43.2046} {Phys. Rev. A {\bf 43},  2046
   (1991)}.

\bibitem{Srednicki1994-50}
M. Srednicki, {\em Chaos and quantum thermalization},
  \href{https://doi.org/10.1103/PhysRevE.50.888} {Phys. Rev. E {\bf 50},  888
  (1994)}.

\bibitem{Rigol2008-452}
M. Rigol, V. Dunjko, and M. Olshanii, {\em Thermalization and its mechanism for
  generic isolated quantum systems}, \href{https://doi.org/10.1038/nature06838}
  {Nature {\bf 452},  854  (2008)}.

\bibitem{Biroli2010-105}
G. Biroli, C. Kollath, and A.~M. L\"auchli, {\em Effect of rare fluctuations on
  the thermalization of isolated quantum systems},
  \href{https://doi.org/10.1103/PhysRevLett.105.250401} {Phys. Rev. Lett. {\bf
  105},  250401  (2010)}.

\bibitem{Beugeling2014-89}
W. Beugeling, R. Moessner, and M. Haque, {\em Finite-size scaling of eigenstate
  thermalization}, \href{https://doi.org/10.1103/PhysRevE.89.042112} {Phys.
  Rev. E {\bf 89},  042112  (2014)}.

\bibitem{Kim2014-90}
H. Kim, T.~N. Ikeda, and D.~A. Huse, {\em Testing whether all eigenstates obey
  the eigenstate thermalization hypothesis},
  \href{https://doi.org/10.1103/PhysRevE.90.052105} {Phys. Rev. E {\bf 90},
  052105  (2014)}.

\bibitem{Alba2015-91}
V. Alba, {\em Eigenstate thermalization hypothesis and integrability in quantum
  spin chains}, \href{https://doi.org/10.1103/PhysRevB.91.155123} {Phys. Rev. B
  {\bf 91},  155123  (2015)}.

\bibitem{Lai2015-91}
H.-H. Lai and K. Yang, {\em Entanglement entropy scaling laws and eigenstate
  typicality in free fermion systems},
  \href{https://doi.org/10.1103/PhysRevB.91.081110} {Phys. Rev. B {\bf 91},
  081110(R)  (2015)}.

\bibitem{Dymarsky2018-97}
A. Dymarsky, N. Lashkari, and H. Liu, {\em Subsystem eigenstate thermalization
  hypothesis}, \href{https://doi.org/10.1103/PhysRevE.97.012140} {Phys. Rev. E
  {\bf 97},  012140  (2018)}.

\bibitem{Deutsch2018-81}
J.~M. Deutsch, {\em Eigenstate thermalization hypothesis},
  \href{https://doi.org/10.1088/1361-6633/aac9f1} {Rep. Prog. Phys. {\bf 81},
  082001  (2018)}.

\bibitem{Yoshizawa2018-120}
T. Yoshizawa, E. Iyoda, and T. Sagawa, {\em Numerical large deviation analysis
  of the eigenstate thermalization hypothesis},
  \href{https://doi.org/10.1103/PhysRevLett.120.200604} {Phys. Rev. Lett. {\bf
  120},  200604  (2018)}.

\bibitem{Barthel2019_12}
T. Barthel and Q. Miao, {\em Scaling functions for eigenstate entanglement
  crossovers in harmonic lattices},
  \href{https://doi.org/10.1103/PhysRevA.104.022414} {Phys. Rev. A {\bf 104},
  022414  (2021)}.

\bibitem{Mori2016_09}
T. Mori, {\em Weak eigenstate thermalization with large deviation bound},
  \href{http://arxiv.org/abs/1609.09776} {arXiv:1609.09776  (2016)}.

\bibitem{Laflorencie2004-645}
N. Laflorencie and D. Poilblanc,  in {\em Quantum Magnetism}, Vol.~645 of {\em
  Lecture Notes in Physics}, edited by U. Schollw\"ock, J. Richter, D.~J.~J.
  Farnell, and R.~F. Bishop (Springer, Berlin, 2004), pp.\ 227--252.

\bibitem{Sandvik2010-1297}
A.~W. Sandvik, {\em Computational studies of quantum spin systems},
  \href{https://doi.org/10.1063/1.3518900} {AIP Conf. Proc. {\bf 1297},  135
  (2010)}.

\bibitem{Bethe1931}
H.~A. Bethe, {\em Zur {T}heorie der {M}etalle. I. {E}igenwerte und
  {E}igenfunktionen der linearen {A}tomkette},
  \href{https://doi.org/10.1007/BF01341708} {Z. Phys. {\bf 71},  205  (1931)}.

\bibitem{Korepin1993}
V. Korepin, N. Bogoliubov, and A. Izergin, {\em Quantum Inverse Scattering
  Method and Correlation Functions} (Cambridge University Press, Cambridge,
  1993).

\bibitem{Nomura1994-27}
K. Nomura and K. Okamoto, {\em Critical properties of S= 1/2 antiferromagnetic
  {XXZ} chain with next-nearest-neighbour interactions},
  \href{https://doi.org/10.1088/0305-4470/27/17/012} {J. Phys. A: Math. Gen.
  {\bf 27},  5773  (1994)}.

\bibitem{Somma2001-64}
R.~D. Somma and A.~A. Aligia, {\em Phase diagram of the XXZ chain with
  next-nearest-neighbor interactions},
  \href{https://doi.org/10.1103/PhysRevB.64.024410} {Phys. Rev. B {\bf 64},
  024410  (2001)}.

\bibitem{Belavin1984-241}
A.~A. Belavin, A.~M. Polyakov, and A.~B. Zamolodchikov, {\em Infinite conformal
  symmetry in two-dimensional quantum field theory},
  \href{https://doi.org/10.1016/0550-3213(84)90052-X} {Nucl. Phys. B {\bf 241},
   333  (1984)}.

\bibitem{Francesco1997}
P. {Di Francesco}, P. Mathieu, and D. Senechal, {\em Conformal Field Theory}
  (Springer, New York, 1997).

\bibitem{Polchinski1988-383}
J. Polchinski, {\em Scale and conformal invariance in quantum field theory},
  \href{https://doi.org/10.1016/0550-3213(88)90179-4} {Nucl. Phys. B {\bf 303},
   226   (1988)}.

\bibitem{Korepin2004-92}
V.~E. Korepin, {\em Universality of entropy scaling in one dimensional gapless
  models}, \href{https://doi.org/10.1103/PhysRevLett.92.096402} {Phys. Rev.
  Lett. {\bf 92},  096402  (2004)}.

\bibitem{Ikeda2011-84}
T.~N. Ikeda, Y. Watanabe, and M. Ueda, {\em Eigenstate randomization
  hypothesis: Why does the long-time average equal the microcanonical
  average?}, \href{https://doi.org/10.1103/PhysRevE.84.021130} {Phys. Rev. E
  {\bf 84},  021130  (2011)}.

\bibitem{Ikeda2013-87}
T.~N. Ikeda, Y. Watanabe, and M. Ueda, {\em Finite-size scaling analysis of the
  eigenstate thermalization hypothesis in a one-dimensional interacting Bose
  gas}, \href{https://doi.org/10.1103/PhysRevE.87.012125} {Phys. Rev. E {\bf
  87},  012125  (2013)}.

\bibitem{Peschel1981-43}
I. Peschel and V.~J. Emery, {\em Calculation of spin correlations in
  two-dimensional Ising systems from one-dimensional kinetic models},
  \href{https://doi.org/10.1007/BF01297524} {Z. Phys. B {\bf 43},  241
  (1981)}.

\bibitem{Sachdev2011}
S. Sachdev, {\em Quantum Phase Transitions}, 2nd ed. (Cambridge University
  Press, Cambridge, UK, 2011).

\bibitem{Fisher1980-44}
M.~E. Fisher and W. Selke, {\em Infinitely many commensurate phases in a simple
  Ising model}, \href{https://doi.org/10.1103/PhysRevLett.44.1502} {Phys. Rev.
  Lett. {\bf 44},  1502  (1980)}.

\bibitem{Rujan1981-24}
P. Ruj\'an, {\em Critical behavior of two-dimensional models with spatially
  modulated phases: Analytic results},
  \href{https://doi.org/10.1103/PhysRevB.24.6620} {Phys. Rev. B {\bf 24},  6620
   (1981)}.

\bibitem{Bak1982-45}
P. Bak, {\em Commensurate phases, incommensurate phases and the devil's
  staircase}, \href{https://doi.org/10.1088/0034-4885/45/6/001} {Rep. Prog.
  Phys. {\bf 45},  587  (1982)}.

\bibitem{Selke1988-170}
W. Selke, {\em The ANNNI model -- Theoretical analysis and experimental
  application}, \href{https://doi.org/10.1016/0370-1573(88)90140-8} {Phys. Rep.
  {\bf 170},  213   (1988)}.

\bibitem{Allen2001-34}
D. Allen, P. Azaria, and P. Lecheminant, {\em A two-leg quantum Ising ladder: a
  bosonization study of the {ANNNI} model},
  \href{https://doi.org/10.1088/0305-4470/34/21/101} {J. Phys. A {\bf 34},
  L305  (2001)}.

\bibitem{Beccaria2006-73}
M. Beccaria, M. Campostrini, and A. Feo, {\em Density-matrix
  renormalization-group study of the disorder line in the quantum axial
  next-nearest-neighbor Ising model},
  \href{https://doi.org/10.1103/PhysRevB.73.052402} {Phys. Rev. B {\bf 73},
  052402  (2006)}.

\bibitem{Beccaria2007-76}
M. Beccaria, M. Campostrini, and A. Feo, {\em Evidence for a floating phase of
  the transverse ANNNI model at high frustration},
  \href{https://doi.org/10.1103/PhysRevB.76.094410} {Phys. Rev. B {\bf 76},
  094410  (2007)}.

\bibitem{Sela2011-84}
E. Sela and R.~G. Pereira, {\em Orbital multicriticality in spin-gapped
  quasi-one-dimensional antiferromagnets},
  \href{https://doi.org/10.1103/PhysRevB.84.014407} {Phys. Rev. B {\bf 84},
  014407  (2011)}.

\bibitem{Dutta2015}
A. Dutta, G. Aeppli, B. Chakrabarti, U. Divakaran, T. Rosenbaum, and D. Sen,
  {\em Quantum Phase Transitions in Transverse Field Spin Models} (Cambridge
  University Press, Cambridge, 2015).

\bibitem{Fradkin2013}
E. Fradkin, {\em Field Theories of Condensed Matter Physics} (Cambridge
  University Press, Cambridge, 2013).

\bibitem{Hassler2012-14}
F. Hassler and D. Schuricht, {\em Strongly interacting Majorana modes in an
  array of Josephson junctions},
  \href{https://doi.org/10.1088/1367-2630/14/12/125018} {New J. Phys. {\bf 14},
   125018  (2012)}.

\bibitem{Cole2017-95}
R. Cole, F. Pollmann, and J.~J. Betouras, {\em Entanglement scaling and spatial
  correlations of the transverse-field Ising model with perturbations},
  \href{https://doi.org/10.1103/PhysRevB.95.214410} {Phys. Rev. B {\bf 95},
  214410  (2017)}.

\bibitem{Mahyaeh2020-101}
I. Mahyaeh and E. Ardonne, {\em Study of the phase diagram of the
  Kitaev-Hubbard chain}, \href{https://doi.org/10.1103/PhysRevB.101.085125}
  {Phys. Rev. B {\bf 101},  085125  (2020)}.

\bibitem{Jordan1928}
P. Jordan and E. Wigner, {\em About the Pauli exclusion principle},
  \href{https://doi.org/10.1007/BF01331938} {Z. Phys. {\bf 47},  631  (1928)}.

\end{thebibliography}
\end{document}